\documentclass[ aps,  prl, twocolumn,  showpacs]{revtex4-1}
\usepackage{epstopdf}
\usepackage{placeins}
\usepackage[normalem]{ulem}
\usepackage{ mathrsfs }
\usepackage{amsfonts }
\usepackage{color,soul}
\usepackage{amsmath}    % need for subequations
\usepackage{graphicx}   % need for figures
\usepackage{color}      % use if color is used in text
\usepackage[normalem]{ulem}
\usepackage{siunitx}
\usepackage[]{units}
\DeclareSymbolFont{rsfso}{U}{rsfso}{m}{n}
\usepackage[T1]{fontenc}

\begin{document}
	\title{Mechanics of a granular \textit{skin}}
	\author{Somnath Karmakar,  Anit Sane,  S. Bhattacharya  and Shankar Ghosh}
	\address{Department of Condensed Matter Physics and Materials Science,  Tata Institute of Fundamental Research,  Homi Bhabha Road,  Mumbai 400005,  India} 
	\date{\today}
	\begin{abstract}
		Magic Sand, a hydrophobic toy granular material, is widely used in popular science instructions because of its non-intuitive mechanical  properties. A detailed study of the  failure  of an underwater column of magic sand shows that these properties can be traced to a single phenomenon: the system self-generates a cohesive \textit{skin} that encapsulates the material inside. The \textit{skin},  consists of pinned air-water-grain interfaces, shows multi-scale mechanical properties: they range from contact-line dynamics in the intra-grain roughness scale, plastic flow at the grain scale, all the way to the sample-scale mechanical responses.  With decreasing rigidity of the \textit{skin}, the failure mode transforms from brittle to ductile (both of which are collective in nature) to a complete disintegration at the single grain scale.
	\end{abstract}
	
	\maketitle

	\section{Introduction}

	Dry granular systems are known to exhibit both liquid and solid-like properties.  While there are many instances where a  granular assembly  exhibits  liquid-like properties  ~\cite{amarouchene_capillarylike_2008, cheng_collective_2007, forterre_flows_2008, cheng_towards_2008},   examples  of granular systems illustrating  solid-like properties are rare with one exception:   the  typical conical sandpile shape ~\cite{deGennes1999}. To craft more complex shapes,  e.g.,  sand art,  it is necessary to add additional constraints.  These constraints can be added in the bulk or at the boundary. Adding a small quantity of water to the sand introduces constraints in the bulk in the form of  capillary bridges ~\cite{strauch_wet_2012, mitarai_wet_2006, soulie_influence_2006,herminghaus_dynamics_2005}. Conversely,    encapsulating  dry grains in  a container  is an example of constraints that are applied at the boundary ~\cite{gutierrez_silo_2015}. 
	
	In  general,  encapsulation requires isolating materials from their surroundings and it is achieved  either  by (i) introducing  another material in the interfacial region or  (ii) facilitating processes at the interface that create  self-encapsulation.  Examples of the former  range from the simple instance of a bag or a silo containing cereal grains to more complex examples of thin polymer films  wrapping liquid droplets ~\cite{py2007capillary,paulsen2015optimal}, texturing of  liquid droplets (`liquid marbles') by attaching hydrophobic powder to its surface ~\cite{aussillous2001liquid} and stabilizing  emulsions by particles or surfactants ~\cite{binks_pickering_2001,dinsmore_colloidosomes:_2002}.  In contrast,  a liquid that  oxidizes on contact with atmosphere  to  develop  a stress bearing  skin   is an example of self-encapsulation ~\cite{dickey2008eutectic}.
	
	In this paper we study the self-encapsulation of a granular system consisting of hydrophobic sand  ~\cite{aussillous2006properties, abkarian2013gravity, subramaniam2006mechanics, MaryamPakpour2012Sandcastle} under water. This system self-generates a \textit{skin}  which  encapsulates    dry hydrophobic sand-grains  and stabilizes the trapped air (bubble) against the force of buoyancy. Removal of trapped air breaks down this encapsulation. The paper also explores the mechanical properties  of this system at  multiple scales: from the  pinning of the three phase contact line at the roughness scale of the particle, plastic flow at the grain-scale, to sample-spanning mechanical responses.  It may be noted here that while hydrophilic granular systems -- both dry and wet -- are widely studied ~\cite{deGennes1999, hornbaker1997keeps}, hydrophobic sand grains submerged in a non-wetting liquid like water remain largely unexplored ~\cite{aussillous2006properties, abkarian2013gravity, subramaniam2006mechanics, MaryamPakpour2012Sandcastle} even though such systems are of practical importance, especially in pharmaceutical, food  and petroleum industries   where newer encapsulation strategies are in great demand ~\cite{Saleh2007323}.

	Mechanical response   of a system is usually  described in a small neighborhood of a reference state.  For a dry granular assembly a `state' of a system is described in terms of the center of mass  of the grains and the forces  among them. For values of  strain smaller than $ 10^{-5} $,   dry granular assembles   show elastic response, i.e.,  the reference  state can be  restored by setting the  applied forces  to zero (see p 92 of ~\cite{andreotti2013granular}).  This elastic response comes from the reversible deformation of the region of contact   between the grains.  However,  to describe  the 
	state of the hydrophobic   sand immersed in water  the information  regarding the center of mass of the grains is not sufficient. We need to augment it with additional information about  the detailed layout of the three phase contact lines and the local contact angles of the \textit{skin}.   The  three  phase contact lines formed at the  grain-water-air interface are immobilized (pinned) by defects present in the system \cite{moulinet2002roughness,quere2005non,quere2003slippy}.    Any additional deformation of the \textit{skin} causes the contact angles to change from their reference values. This  generates  restoring forces in the system, i.e., if the contact angle changes from $ \theta  $ to $ \theta^d $ the restoring force  per unit length is $ \simeq\gamma(\cos \theta^{d} -\cos \theta) $, here $\gamma $ is the  surface energy  at the water-air interface. The pinning and depinning of these contact lines determine the mechanical properties of the material and  its mode of failure. Hence, the present study is an  example of the more general problem of studying statics and dynamics of systems with elastic interfaces  in a random environment \cite{miguel2006jamming}.

	%%%%%%%%%%%%%%%%%%%%%%%%%%%%%%%%%%%%%%%%%%%%%%%%%%%%%%%%%%%%%%%%%%%%%%%% 
	%%%%%%%%%%%%%%%%%%  Figure introduction:  %%%%%%%%%%%%%%%%%%%%%%%%%%%%%% 
	%%%%%%%%%%%%%%%%%%%%%%%%%%%%%%%%%%%%%%%%%%%%%%%%%%%%%%%%%%%%%%%%%%%%%%%
	
	\begin{figure*}[t]
		\centering {\includegraphics[width=1\linewidth]{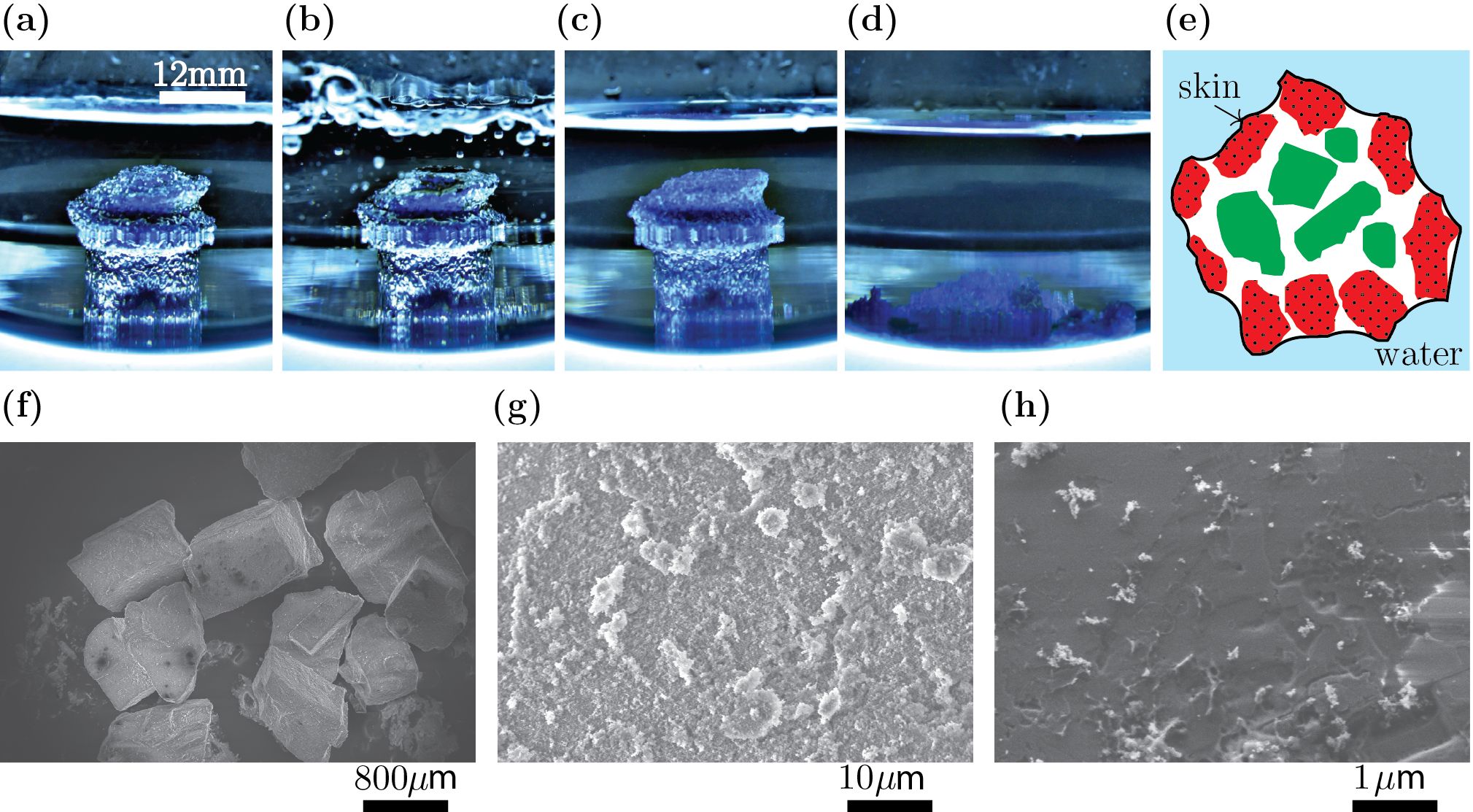}}
		\caption{ Image (a) shows an underwater lump-like rigid structure when magic sand is poured into water. The lump has luster on its surfaces caused by total internal reflection of light at the air-liquid interfaces. When air is gradually removed by pumping (see (b) the lump loses its luster (as seen in (c). This luster-less lump, when mechanically perturbed,  disintegrates into a shallow flat bed (see (d). A cross-sectional top-view of the lump in (a) which is schematically shown in (e), captures the liquid shapes at the `grain-water-air' interfaces. The grains that reside on the lustrous \textit{skin} are marked in red (with dots) and the dry grains in the interior are marked in green; the water and air phases are coloured  in blue and white, respectively. Note, that the \textit{skin} supports the weight of the interior dry grains  and is hence under tension.	The panels (f) and (g) show the scanning electron micrographs of the sand grains for two different values of magnification.  The grain surfaces are decorated with micron-sized hydrophobic patches. Washing hydrophobic sand in acetone removes the surface patches and makes the material hydrophilic. The micrograph of the acetone-washed sand grain  is shown in panel (h).
		}
		\label{fig:intro}
	\end{figure*}
	%%%%%%%%%%%%%%%%%%%%%%%%%%%%%%%%%%%%%%%%%%%%%%%%%%%%%%%%%%%%%% 

	\section{The hydrophobic grains and their wetting properties}\label{Materials and its properties}

	As hydrophobic granular material, we use polyhedral shaped `Magic Sand' grains sourced from Education Innovation Inc. (USA). They are made by coating polyhedral shaped sand grains with a hydrophobic material.   Alternatively,  these  hydrophobic  particles can be made  in the laboratory   by coating    similar sand particles  with    FluoroPel PFC from Cytonix LLC silica ~\cite{varshney_amorphous_2012}.

	When  grains of magic sand are freely poured into water, they spontaneously form a cohesive lump: a typical example of which is shown in Fig.\ref{fig:intro}(a). The immediately evident features   are: (i) the system retains its shape and (ii) the outer surface of the lump has a luster that originates from the  total internal reflections of light at the pinned water-air interface ~\cite{Davis_1955,Marston_1979}. On degassing of the system achieved by creating a partial vacuum over the liquid (Fig.\ref{fig:intro}(b), the lump begins to lose its lustre (Fig.\ref{fig:intro}(c) and when this luster-free system is mechanically perturbed, it slumps to form a flat sand-bed inside water (Fig.\ref{fig:intro}(d). A cross-sectional top-view of the granular lump in Fig.\ref{fig:intro}(a) is schematically shown in Fig.\ref{fig:intro}(e). The grains (red coloured with dots) on the boundary  together with the pinned water-interface constitute an encapsulating granular \textit{skin} which provides the submerged lump its structural integrity. The  force of buoyancy of the trapped air and the weight of  dry sand grains  in the interior (colored with green) exerts stress on the \textit{skin} causing it to be under tension.

	\begin{figure}[t]
		\centering
		\includegraphics[width=1\linewidth]{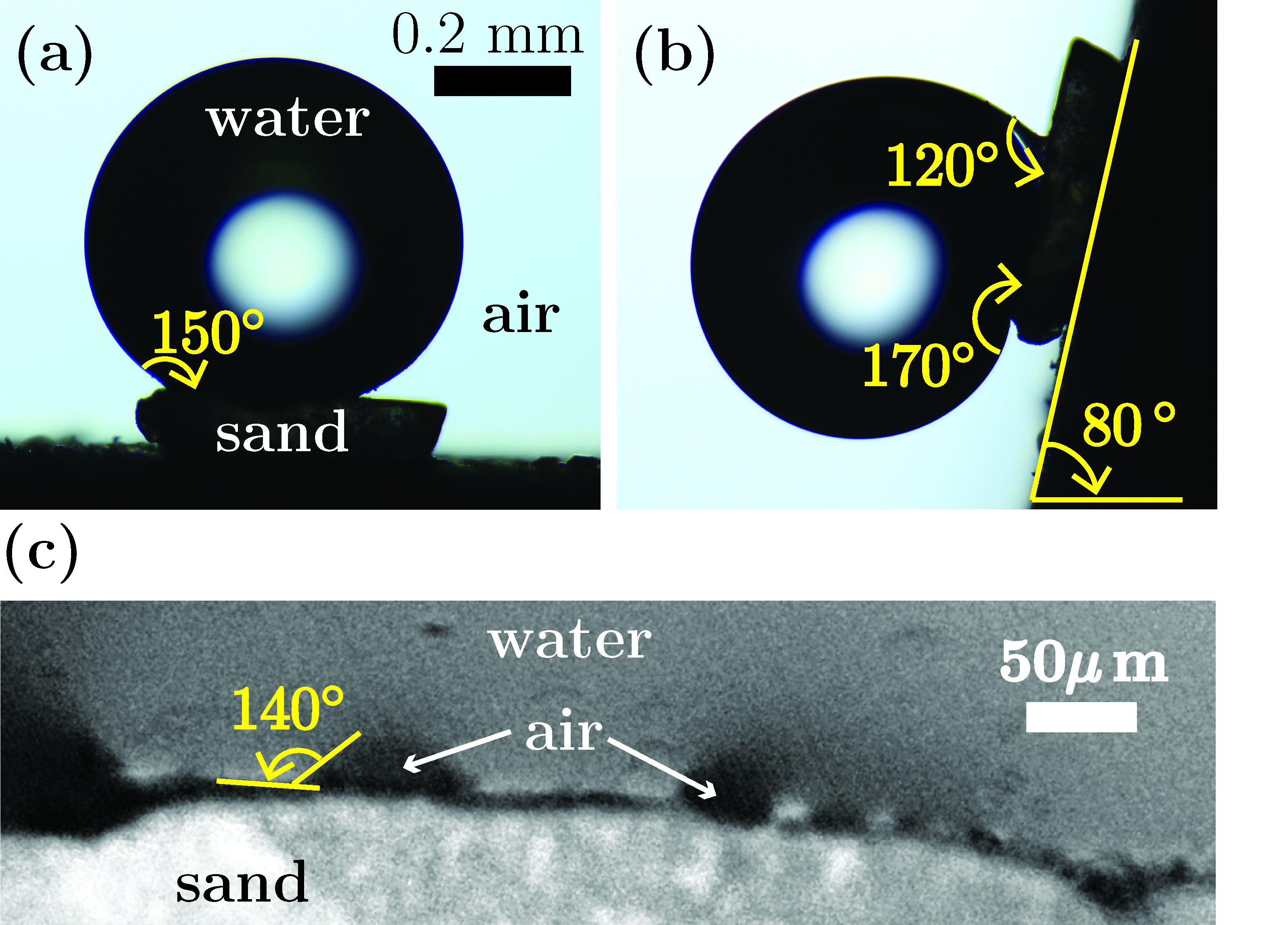}
		\caption{ (a) Equilibrium contact angle of $ \approx 150^{\circ} $ obtained for a water-drop ($ \approx \, 50 \, \mu L $) placed on a single hydrophobic grain surface. (b) Advancing and receding contact angles are measured to be $ \approx 170^{\circ} $ and $  120^{\circ} $ respectively, for the same water droplet in a tilted configuration. (c)  Existence of isolated air bubbles on the topographic grain surface viewed at a larger magnification. }
		\label{fig:contact_angle}
	\end{figure}

	The scanning electron micrographs  of the grain surface are shown in Fig.\ref{fig:intro}(f) and  (g), for two different values of magnification. The micron-sized rough patches (see Fig.\ref{fig:intro}(g) which come from the  hydrophobic coating on the particle surface are crucial for the observed wetting properties of these sand grains ~\cite{quere2005non}.  The equilibrium contact angle is measured to be about $ 150^{\circ} $ for a $ \approx 50 \, \mu L $ water drop placed on a  single hydrophobic grain (see Fig.\ref{fig:contact_angle}(a). For the same droplet in a tilted configuration, the  advancing and receding contact angles
	are found to be $ \approx 170^\circ $ and  $ \approx 120^\circ $ respectively, i.e., the contact angle hysteresis is $ \approx 50 ^{\circ} $; see 
	Fig.\ref{fig:contact_angle}(b).  
	The higher magnification image of water-grain interface in Fig.\ref{fig:contact_angle}(c) shows  presence of  small isolated air bubbles trapped between the grain and the water which have similar equilibrium contact angles  $ (\approx 140^\circ) $, i.e., the surrounding water is in partial contact with the hydrophobic sand grains which creates a Wenzel type wetting scenario \cite{quere2003slippy,quere2005non}. Washing these  grains  in acetone removes the hydrophobic coating and exposes the underlying smooth  hydrophilic  surface (Fig.\ref{fig:intro}(h).

	%%%%%%%%%%%%%%%%%%%%%%%%%%%%%%%%%%%%%%%%%%%%%%%%%%%%%%%%%%%%%%%%%%%%%%%% 
	%%%%%%%%%%%%%%%%%%  Figure modulation:  %%%%%%%%%%%%%%%%%%%%%%%%%%%%%%% 
	%%%%%%%%%%%%%%%%%%%%%%%%%%%%%%%%%%%%%%%%%%%%%%%%%%%%%%%%%%%%%%%%%%%%%%%% 
	
	\begin{figure*}[t]
		\centering {\includegraphics[width=1.00\linewidth]{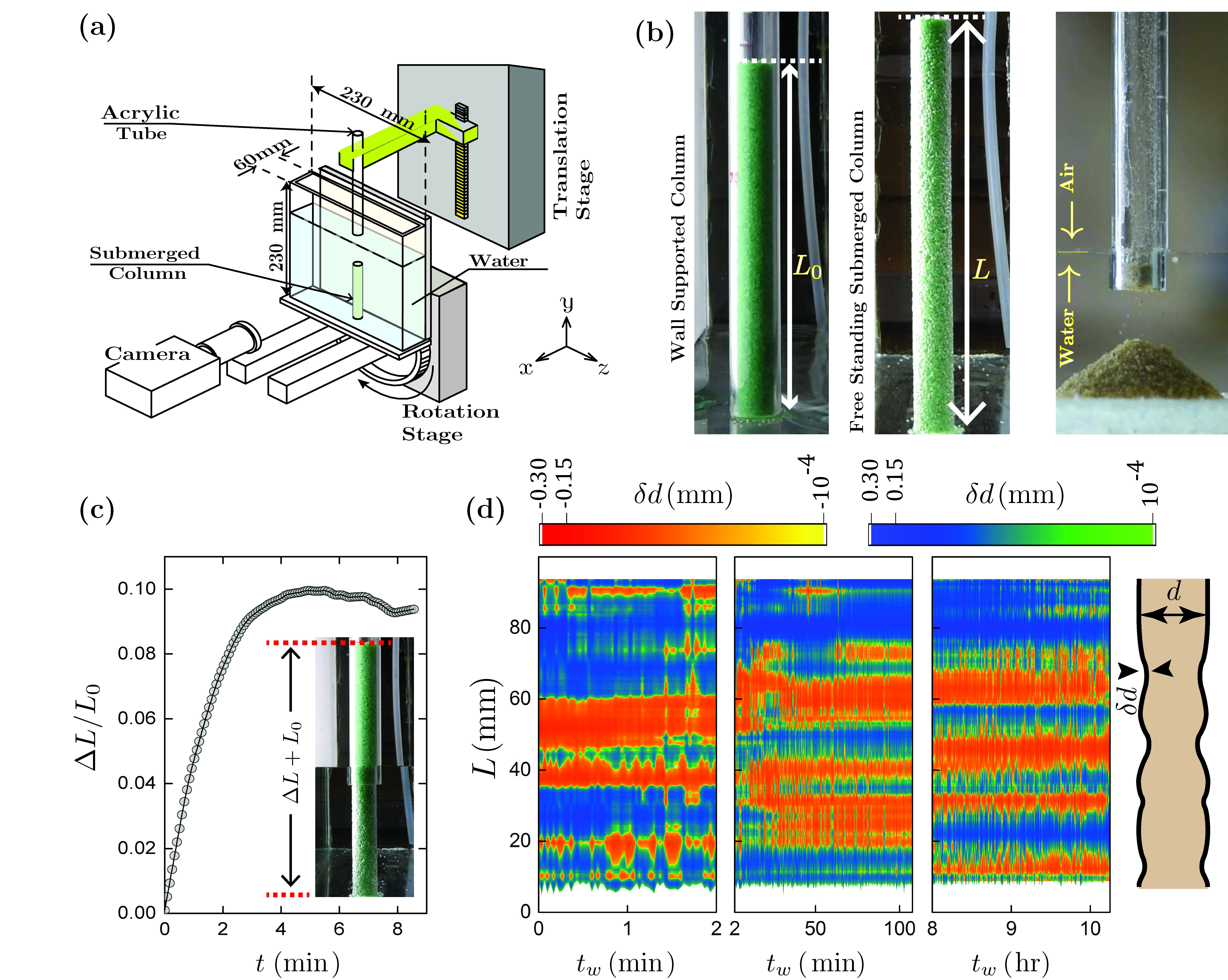}}
		\caption{ (a) Schematic of the experimental setup to study the mechanical response of sand columns in the presence of body forces. The vertical water-filled glass container, mounted on a rotary stage, is gradually tilted to apply body forces to the column. 
			(b) Images of a \textit{wall-supported} (left panel)  and a submerged hydrophobic sand columns (middle panel). The right panel contrasts the response of hydrophilic sand in water  that  slumps to form a sand pile. (c) Axial strain $\Delta L/ L_0$ as a function of time $t$ while the column is being made; Inset: column preparation at an instance time.  (d) The contour plots show time evolution of the  changes in the diameter $\delta d$  from its mean value for  the submerged column ($d=11\, \mathrm{mm}$,  $a = 500\, \mathrm{\mu m}$) along its vertical length $L$. The blue and the orange bands correspond to regions with positive ($ \delta d>0 $) and negative ($ \delta d<0 $) modulations of the diameter, respectively. The right-most panel shows a schematic representation of the peristaltic modulations created by the wrinkles on the \textit{skin}. 
		}
		\label{fig:Fig_modulation}
	\end{figure*}
	%%%%%%%%%%%%%%%%%%%%%%%%%%%%%%%%%%%%%%%%%%%%%%%%%%%%%%%%%%%%%% 

	%%%%%%%%%%%%%%%%%%%%%%%%%%%%%%%%%%%%%%%%%%%%%%%%%%%%%%%%%%%%%% 

	\section{Construction of a submerged free-standing column}
	
	In order to obtain a more comprehensive understanding of the mechanical properties of the system, we choose a relatively simple geometry which is a free-standing submerged cylindrical column built with magic-sand, using the following experimental steps.  At First, by selective sieving we obtain grains of different mean sizes $a$, ranging from $250 \, \mathrm{\mu m}$ to $2 \, \mathrm{m m}$. Second, an acrylic tube  of inner diameter $ d $ is made to  rest vertically on the flat base of an empty rectangular glass container. The grains are thereafter gently poured into the tube upto a desired height $ L_0 $. This forms a \textit{wall-supported} granular column shown in  the left panel of Fig.\ref{fig:Fig_modulation}(b). The typical packing fraction $ \phi$ of the grain assembly is about $ 0.53 $. 
	Third, and in a key step, the acrylic tube is gradually pulled out while simultaneously filling the glass container with water. Consequently, the hydrostatic pressure partially compensates for the over-pressure due to weight of the grains. This unique protocol prevents grains from  clogging
	%; this  is seen  in the left panel of supplementary movie M1 \cite{M1}. The schematic drawing of our experimental setup is given in Fig.\ref{fig:Fig_modulation}(a).

	Particles which are in constant physical contact  with the wall of the acrylic tube are immobilized by the frictional interactions.    As the  tube  is withdrawn,  it drags along  particles that are   in contact with it 
	%(see left panel of  movie M1 in the Supplemental Material \cite{M1}).
	  This exposes the next layer of particles to  water,  on which the  \textit{skin} forms. The resulting particle rearrangements reduce the diameter by $ \approx 1\% $ and increase the height of the column  which is now partially submerged in water by $ \Delta L $. The radial   contraction happens  within the first few seconds and does not evolve appreciably over time. However, the axial strain $\Delta L/L_o$  keeps increasing in time; see Fig.\ref{fig:Fig_modulation}(c). As the tube is pulled out,  the number of  sand grains which are in contact with it  decreases.  This reduces the extent of particle rearrangements.  As a result, the rate of change of $ \Delta L $ decreases with time.  	The fully submerged column  has a length $ L $ which is about $ \approx 9 \% $ larger than  the initial  length $ L_0 $ of the wall-supported column. Hence, the volume fraction of the submerged column is about $ 7 \% $ lower than its value for the wall-supported case.   This free-standing submerged column  is shown in the middle panel of Fig.\ref{fig:Fig_modulation}(b). The  column is anchored via capillary forces with the hydrophobic substrate at the bottom and is slightly flared  close to the anchoring region.  Under an identical protocol, hydrophilic beach sand instead slumps to form a canonical sandpile ~\cite{rondon2011granular}.
	  %; see the right panel of Fig.\ref{fig:Fig_modulation}(b) and the  movie M1 in the  Supplemental Material \cite{M1}. 

	\section{Mechanical properties of the column}
	
	The submerged column can be thought of as a thin cylindrical shell (made from the \textit{skin}) containing non cohesive dry grains in its interior. The  \textit{skin}  is about a particle diameter thick and it comprises of the  sand grains on the outer surface of the column and the pinned three phase contact line.  Using a  Du-No{\"u}y ring based technique ~\cite{du1925interfacial}    we measure the   interfacial energy of the \textit{skin} to  be about \SI{0.05}{N/m}.  
	%\subsection{Pinning of the contact line at the micron scale of the hydrophobic patches}  
	At a microscopic scale  this interfacial energy arises  from the pinning  of the  three phase contact line to the  defects present on the  surface of the grains (see Fig. \ref{fig:contact_angle}(c) \cite{moulinet2002roughness}.   The  cohesive coupling stiffness $ \mathscr{K} $ due to this pinning depends on the size of the defect $ \ell_{dis}$   and the typical separation  $ \zeta $  between them. In the limit of strong but sparsely distributed defects,  $\mathscr{K}=\pi \gamma/(\ln\left(\zeta/\ell_{dis}\right), $ where $ \gamma $ is the  surface energy  at the water-air interface \cite{RevModPhys.57.827}.  Experimentally we observe that  polyhedral grains   with smooth coverage of hydrophobic coating   consistently fail to sustain  submerged columns. This  suggests that of the multiple scales present in the system,  the  micron-sized hydrophobic patches present on the surface of the grains are most effective in pinning the contact line.

	%%%%%%%%%%%%%%%%%%%%%%%%%%%%%%%%%%%%%%%%%%%%%%%%%%%%%%%%%%%%%%%%%%%%%%
	%%%%%%%%%%%%%%%%%%  Figure Curvature:  %%%%%%%%%%%%%%%%%%%%%%%%%%%%%%% 
	
	\begin{figure}[t]
		\centering {\includegraphics[width=0.99\linewidth]{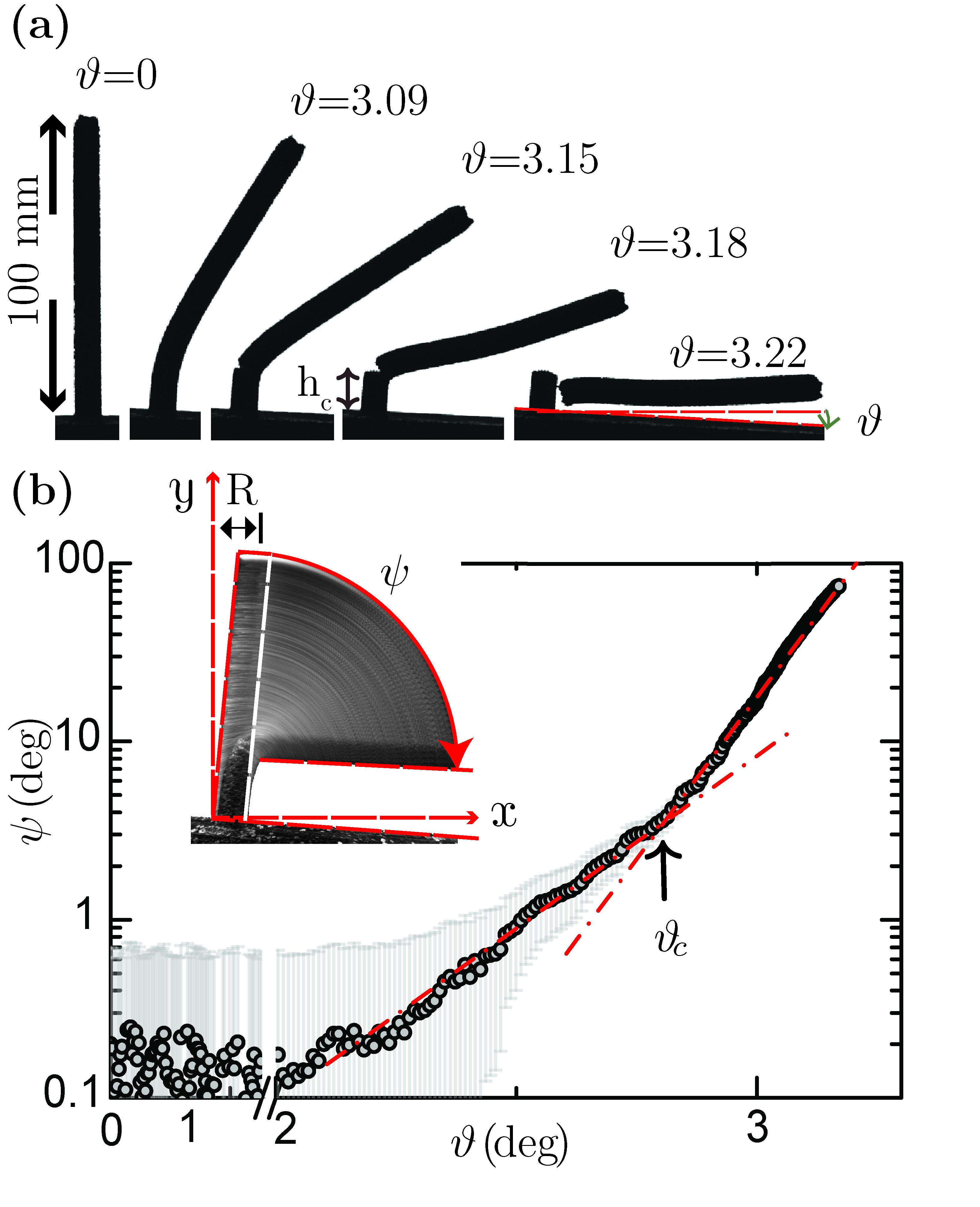}}
		\caption{(a)  The figure shows sequence of images for increasing tilt angle $ \vartheta $ which  captures the  mechanical failure of the submerged column. 
			Note that unlike the case of a  ``falling chimney'' which breaks \textit{backward} in  mid-air
			~\cite{varieschi2003toy,sutton1936concerning}  the submerged column breaks \textit{forward} (with the exception of a  slight catenary shape seen towards the top  for $ \vartheta =3.18 $).  
			(b) Variation of the failure angle $\psi $ on the tilt angle $ \vartheta $ (protocol of continuous tilting). Beyond the critical tilt angle $ \vartheta_c $ the  variation of $ \psi $ with $ \vartheta $ changes its slope on a semi-log plot  (shown by the red colored dash-dotted lines). The inset shows the overlay of the images for falling of the column; where $  R $  is the distance of the mid-point of the top of the column from the point $  (0,L) $  on the $y-$axis. }
		
		\label{fig:Fig_curvature}
	\end{figure}
	%%%%%%%%%%%%%%%%%%%%%%%%%%%%%%%%%%%%%%%%%%%%%%%%%%%%%%%%%%%%%%%%%%%%%%\dfrac{•}{•}
	
	\subsection{Growth of imperfections of the column }
	
	The process of making the column introduces localized geometric imperfections which evolve   slowly in time.  This   is captured by the contour plots of the variation $ \delta d $ of the column diameter from its mean value along its length   over four orders of magnitude in waiting time $ t_w $ (Fig.\ref{fig:Fig_modulation}(d). 
	In Fig.\ref{fig:Fig_modulation}(d)  the  blue regions  correspond to regions with positive,   $ \delta d>0 $, modulation and the orange  regions  correspond to regions with negative,   $ \delta d>0 $, modulation. The maximum $ \delta d $ is found to be around $ 0.3 \, \mbox{mm} $, nearly half-a-particle size $ a $; whereas the width of the modulation in $ d $ along $ L $ is much larger, approximately 40 particle-lengths. The slow temporal evolution of the column's deformation  is indicative of creep in the presence of strong inter-grain friction,  brought about by the competition between the  stresses  exerted on the \textit{skin}  by the  dry grains in the interior \cite{gutierrez_silo_2015} and the interfacial energy density of the \textit{skin}. Henceforth,  we   refer to it as the wrinkling of the \textit{skin}. 
	
	%Even in the limit of random frictional mobilization, this compressive traction force can either reinforce imperfections on the \textit{skin} introduced during  its formation  or generate new  localized imperfections of the \textit{skin} \cite{gutierrez_silo_2015}. 

	\subsection{  An order of magnitude estimate of the elastic constant   }  
	
	In this section we make an order of magnitude estimation  of an effective elastic constant for the column.
	To do so, we  apply increasing body forces to the column  by gradually tilting the vertical  water-filled glass container at a fixed rate; for experimental details see  Fig.\ref{fig:Fig_modulation}(a).
	 %The  movie M2 in the Supplemental Material \cite{M2}  captures the mechanical response of the column and  
	 Fig.\ref{fig:Fig_curvature}(a) displays five representative images of the column for increasing values of the tilt angle $ \vartheta $  made by the base of the container with respect to the horizontal. Images corresponding to $ \vartheta = 3.15 ^{\circ} $ and  $ \vartheta =3.18^{\circ} $ capture the process of the column breaking which resembles  a mode-I type transverse rupture.  As the column breaks from a finite column height $ h_c $,  its upper-part   traces out a distinct arc, $ y\simeq h_c \cos\vartheta+ L_{\psi}\cos \psi \cos\vartheta$ (the inset of Fig.\ref{fig:Fig_curvature}(b).  Here $ \psi $ is the failure angle made by the axis of the falling  part with a line whose slope is $ \cot \vartheta $ and $ L_{\psi} $ is the length of the falling part of the column. The   column for $ \vartheta=3.15$ is about $ 3\% $ longer than  its  length for  $ \vartheta=0$.  The variation of failure angle $ \psi $  with    $ \vartheta $ changes slope at $ \vartheta_c $ in a semi-log plot (see Fig.\ref{fig:Fig_curvature}(b). The large slope beyond $\vartheta_c$ indicates a rapid toppling of the upper part of the column. We use $\vartheta_c$ to mark the onset of mechanical failure.  The broken part of the column sinks as a single object.

	%%%%%%%%%%%%%%%%%%%%%%%%%%%%%%%%%%%%%%%%%%%%%%%%%%%%%%%%%%%%%% 
	%%%%%%%%%%%%%%%%%%  Figure Scaling:  %%%%%%%%%%%%%%%%%%%%%%%%%%%%%%% 
	
	\begin{figure*}[t]
		\centering {\includegraphics[width=1\linewidth]{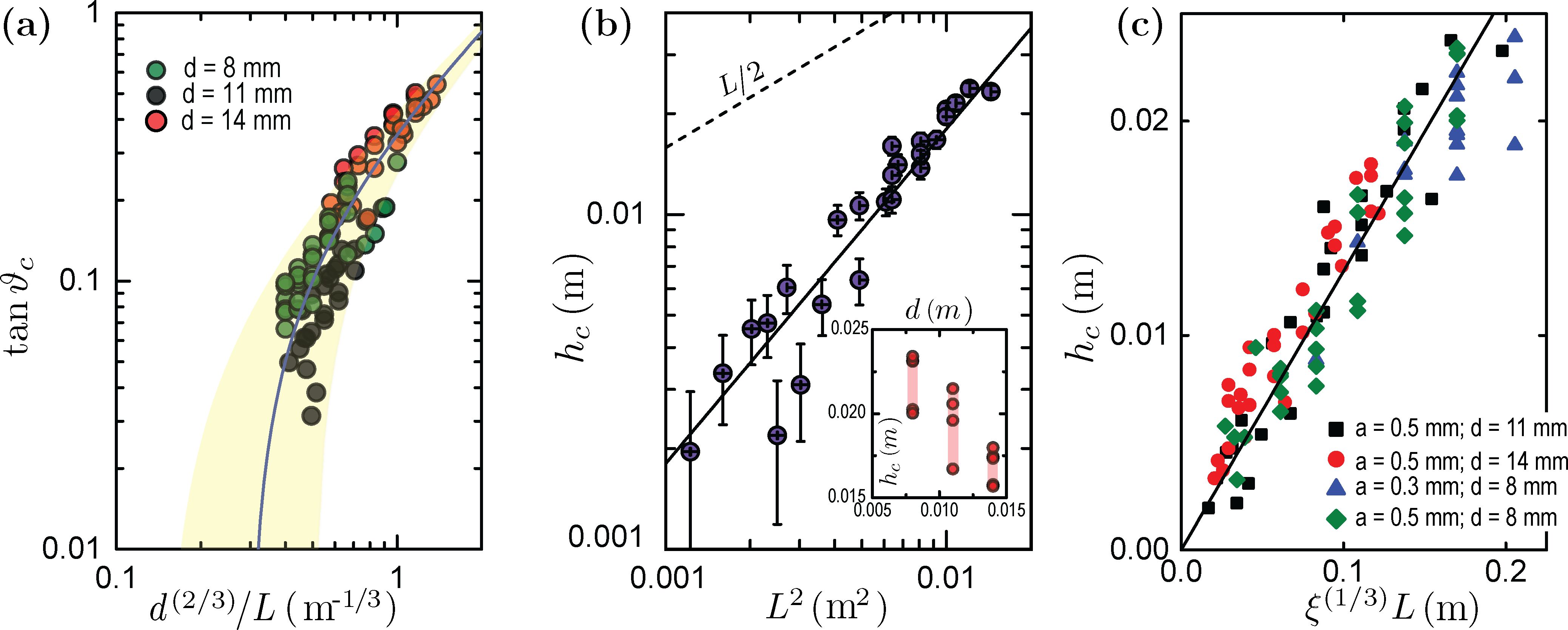}}
		\caption{(a):  Variation of $\tan \vartheta_c $ as a function of $d^{2/3}/L$. The solid  line is a numerical fit for the expression: $\tan \vartheta_c\simeq -c_o + (\frac{\ell_{\vartheta} d^2}{L^3} )^{\frac{1}{3}}$ where  $c_0 \approx 0.15$ and  $\rm \ell_{\vartheta} \approx 0.12 \, m$ are  obtained from fitting. The confidence band of the fitting is marked in yellow. (b)	Variation of $h_{c}$ as a function of $L^2$, obtained for $d=11 \, \mathrm{mm} $ and $a=500 \,\mathrm{\mu m}$. The solid line is a linear fit of $h_c= L^2/\ell_{\beta}$, where $\ell_{\beta}\approx 0.5\, \mathrm{m}$. The inset  shows the  variation of $h_{c}$ as a function of $d$, obtained for  $ L= 100\, \mathrm{mm} $ and $ a= 0.5 \, \mathrm{mm} $.    For a contrast, the dashed line $h_c = L/2$  corresponds to the failure criteria of the chimney's falling for small tilt angles  \cite{varieschi2003toy}.   (c): Variation of $h_{c}$ with $\xi^{1/3}L$ for submerged columns of different $d$'s, $a$'s and $L$'s. The solid line shows the linear dependence.	}
		\label{fig:Fig_scaling_B}
	\end{figure*}

	%\subsection{Mechanical properties of the \textit{skin}} 

	%\subsection{An order of magnitude calculation of the elastic constant for the \textit{skin}}

	One cannot build an arbitrarily long and/or thin column. If the gravitational energy $ \simeq  A(\rho-\rho_f)\phi g L$  of the column exceeds  its bending energy  $  \simeq \frac{YI}{L^2} $, the column will break, here $ Y$ is the effective  Young's Modulus of the column,
	$ A= \frac{1}{4}\pi d^2$  is its cross sectional area,   
	$ I=A^2/2\pi$ is   moment of inertia  of the  column  and $ \rho $ and $\rho_f$ are the densities of sand and water, respectively.     To  estimate  the maximum height $ L_m$  of a column   we equate the gravitational energy to the bending energy to  obtain, 
	\begin{equation*}\label{Eqn1}
	\left(\frac{d^{2/3}}{L_m}\right) =  \left( \frac{(\rho-\rho_f)\phi g}{Y} \right)^{1/3}. 
	\end{equation*}
	% from the following  stability criterion \,~\cite{cox_shape_1998}, i.e., 
	%$ \left(\frac{d^{2/3}}{L_m}\right) \simeq  \left( \frac{(\rho-\rho_f)\phi g}{Y} \right)^{1/3}. $
	At this maximum  height the column breaks in its upright position, i.e., $ \vartheta_c =0 $; this allows us to construct an alternative estimate of $ L_m$ from the experimentally observed linear variation $ \tan \vartheta_c= -c_o + (\frac{\ell_{\vartheta} d^2}{L^3} )^{\frac{1}{3}}$  where $ c_0 \approx 0.15$  and  $ \rm \ell_{\vartheta} \approx$ \SI{0.12}{m} are fitting parameters; see Fig.\ref{fig:Fig_scaling_B}(a). For $ \vartheta_c=0$, we obtain 
	\begin{equation*}\label{Eqn2}
	\left(\frac{d^{2/3}}{L_m}\right)=\left(\frac{c_o^3}{\ell_{\vartheta} } \right)^{1/3}.  
	\end{equation*}
	For $ d = 11\, \mbox{mm} $,  $ L_m = \left(\frac{\ell_{\nu}d^2}{c_o^3}\right)^{\frac{1}{3}}\approx 0.16 \, \mbox{m} $. This estimate of $ L_m $ is consistent with experimental observations that stable $11\, \mbox{mm} $ diameter columns could only be built to the height of $ \approx 0.15\, \mbox{m} $. Columns built greater than this height are unstable and fail in their upright position.
	Equating the above two expressions for $ \left(\frac{d^{2/3}}{L_m}\right)$  we obtain   $  Y  \,\simeq \phi g(\rho-\rho_f)(\ell_{\vartheta}/c_0^3) \approx 10^5 \,\mbox{Pa}$  for the submerged column. This is about two orders of magnitude smaller than that observed for wet sand containing $ 1\% $ water by volume \cite{moller2007shear}.

	\paragraph{Heuristic arguments for calculating $ h_c $:} The falling column  is  visually similar to the  well studied problem of a  ``falling  chimney'' that breaks mid-air due to tilting 
	~\cite{varieschi2003toy,sutton1936concerning}. For the case of a falling chimney  the 
	tensile stress at the leading-edge and the compressive stress at the trailing-edge at small  tilt angles    is maximum at   $\sim L/2 $  (for tilt angles greater than $ 45^{\circ} $ it is about $ L/3 $),  and is usually the point from where the chimney fails \cite{varieschi2003toy}. Though, one expects a similar spatial variation of the  stress  for the submerged column,  when $ L<L_m $, we find the following (i)  the height $ h_c$ at which the column breaks varies quadratically with the column length $ L$; see Fig.\ref{fig:Fig_scaling_B}(b)  and (ii) $ h_c $ decreases with increasing column diameter; see the inset of Fig.\ref{fig:Fig_scaling_B}(b).
	
	From a dimensional argument, for values of $ L $ smaller than $ L_m $,  we propose  that the height $ h_c$  is  related to the column length $  L $  by the following relation $ h_c = f(\xi) L$ where $ \xi=A\phi g (\rho-\rho_f)L^3/YI= 8\phi g (\rho-\rho_f)L^3/Yd^2 $  is the  ratio of the gravitational and the bending energies~\cite{mahadevan1996coiling}. Since $ h_c \propto L^2$, $ f (\xi)$  must vary as $ \xi^\frac{1}{3}$, i.e.,
	\begin{equation*}
	h_c \propto  \xi^{1/3}L.
	\end{equation*}
	A good agreement of this expression with a large set of experimental data, obtained for different lengths and diameters of the submerged column and for various grain sizes, are shown in Fig.\ref{fig:Fig_scaling_B}(c). Here, we have taken the effective Young's modulus of the column  $ Y(= \rm 10^5 \,Pa )$ to be a constant, independent of the grain scale, which implies an approximate validity of a continuum mechanical description of the system in this range of deformation.

	\subsection{The connection between  the  wrinkling of the \textit{skin} and the  critical height $ h_c $    }
	\begin{figure}[b]
		\centering {\includegraphics[width=1\linewidth]{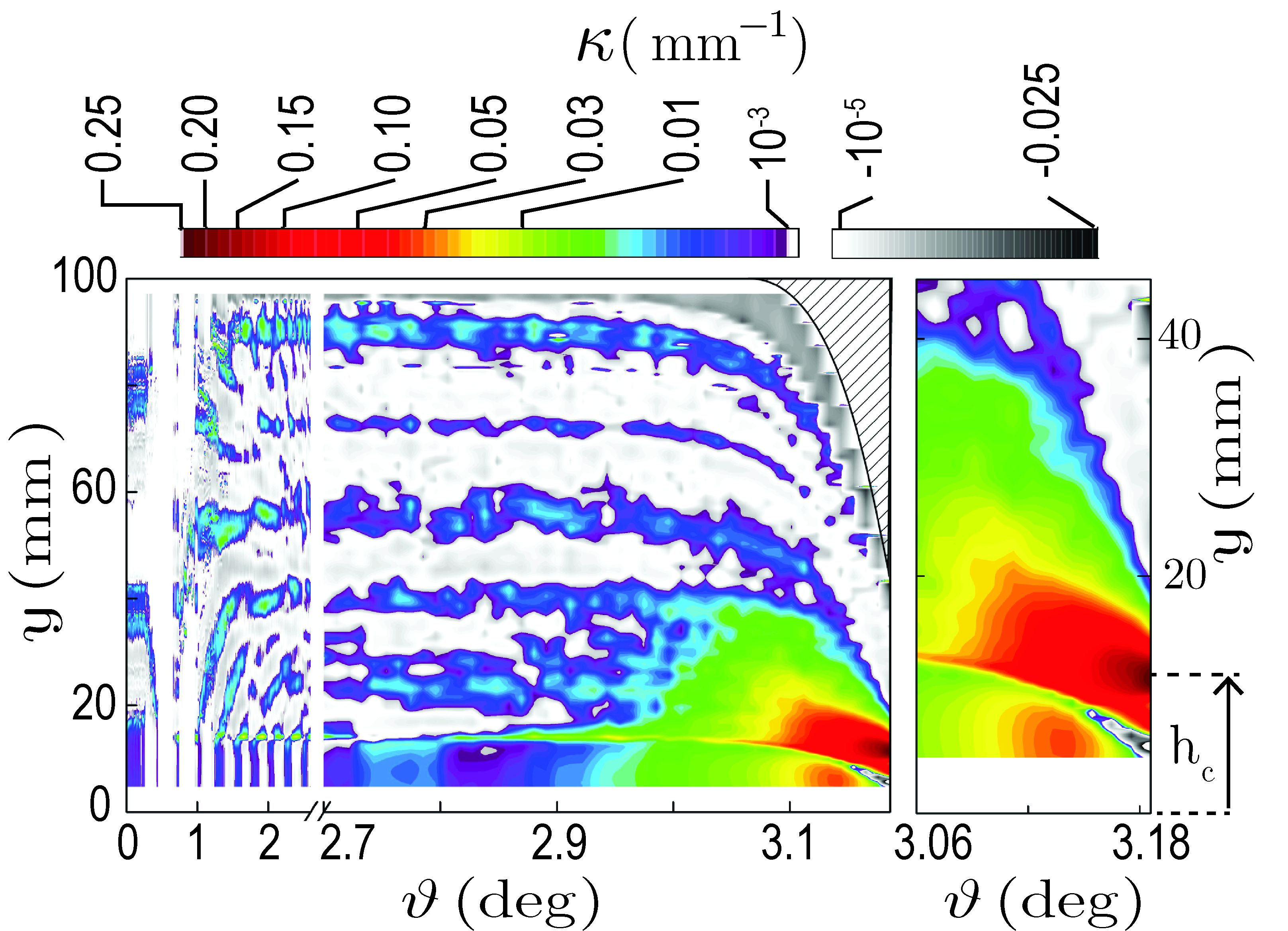}}\\
		\caption{ Left panel: Contour plot of the curvature $ \kappa $ along $ y $-axis for varying $ \vartheta $.  The curvature data is calculated for the advancing (right) edge.    The column can take in any value within the range $ \{0, y_{max}\} $ on the $ y $ axis. When $ \vartheta =0$ the maximum  projected height of the column on the $ y $ axis equals the total length of the column, i.e.,   $ y_{max}=L $.  As the column tilts $y_{max}$ decreases. The  hatched region  corresponds to the values on the $ y  $ axis where $ y>y_{max} $. Right panel: zoomed-in portion of this same plot at close to the failure height $ h_c $.
		}
		\label{fig:Fig_curvature_B}
	\end{figure}
	%%%%%%%%%%%%%%%%%%%%%%%%%%%%%%%%%%%%%%%%%%%%%%%%%%%%%%%%%%%%%%%%%%%%%%%\dfrac{•}{•}

	The  functional dependence of the critical  buckling stress on the geometrical parameters for a cylindrical column is  different from that of  a cylindrical shell. For the column the  critical stress is  proportional to $(d/L)^2 $ \cite{timoshenko1970theory} whereas for a shell it is proportional to $ (a_{\delta}/d)  $ \cite{timoshenko1970theory}. Here  $ a_{\delta} $ is the thickness of the shell. In the present experiments $ (d/L) \approx 0.1 $  and  $(a_{\delta}/d) \approx 0.005  $, where  the shell thickness $ a_{\delta} $ is  of the order of particle diameter $ \approx 0.5 \,  \mbox{mm} $.   Thus, there are two critical stress values; one that corresponds to the buckling of the shell (\textit{skin}) and the other that corresponds to the buckling of the column.   Since  $ (d/L)^2 > (a_{\delta}/d) $, for increasing values of stress,   the buckling of the shell which   causes  the \textit{skin}  to wrinkle, precedes the failure of the column (assuming that the elastic constant of the entire column is greater than or  equal to the elastic constant  of the \textit{skin} alone).

	The \textit{skin} is  an integral  part of the column and its wrinkling   influences the buckling of the column in the following way.  The wrinkles on the \textit{skin} generates geometric imperfections of  the column. These imperfections (i) lowers  the critical stress at which the system fails \cite{hutchinson1970postbuckling} and (ii)  influence  the location from where the system fails.  To study the  role played by  the wrinkles (geometric imperfections of the column) in determining  the height  $ h_c $ from which the column fails we track the curvature $ \kappa $ of the column along its length for the leading edge till the column breaks up into two pieces.

	To calculate $ \kappa $,  the  boundaries of the leading and the trailing edges   are passed  through a low pass filter that suppresses features smaller than a single grain size ~\cite{weinhart_coarse-grained_2013}. These edges  are  detected using an edge detection algorithm based on \cite{trujillo2013accurate}. Fig.\ref{fig:Fig_curvature_B} shows the curvature data for the advancing edge.
	
	The contour plot of the curvature   clearly shows the existence of  bands of high curvature  along the length of the column. These bands correspond to the wrinkling of the \textit{skin}  (Fig.\ref{fig:Fig_modulation}(d).  As the column tilts the lowest wrinkle develops into   the  most prominent imperfection of the column and it acts as a seed from where the failure of the column is initiated.

	The critical height $ h_c $ from where the column breaks   varies  quadratically with the length $ L $ of the column.    Since the location of the lowest wrinkle coincides with $ h_c$,  it is natural to expect the  wavelength associated with the wrinkles to have the same $ L $ dependence as $ h_c $. However, within the framework of shell buckling, the wavelength of  wrinkles is independent of the length of the column \cite{timoshenko1970theory}.  So,   a model based solely on buckling produced by  static loading  is inadequate to describe the experimental scenario.

	We speculate that the observed  length  dependence of $ h_c $ is related to the  perturbations imparted to the column by the mechanical noise associated with the process of tilting. This   gives the loading a certain dynamic character which could in principle
	facilitate  coupling of the  various global length-dependent buckling modes  to the wrinkling induced localized imperfections of the column \cite{hunt2003cylindrical,abramowicz1997transition}.   This conjecture needs to be examined in future  in greater detail.

	%%%%%%%%%%%%%%%%%%%%%%%%%%%%%%%%%%%%%%%%%%%%%%%%%%%%%%%%%%%%%% 

	\subsection{Opening of the wedge between the falling top and the anchored bottom   of the column }
	%%%%%%%%%%%%%%%%%%%%%%%%%%%%%%%%%%%%%%%%%%%%%%%%%%%%%%%%%%%%%% 
	%%%%%%%%%%%%%%%%%%  Figure Scaling:  %%%%%%%%%%%%%%%%%%%%%%%%%%%%%%% 
	
	\begin{figure}[t]
		\centering {\includegraphics[width=1\linewidth]{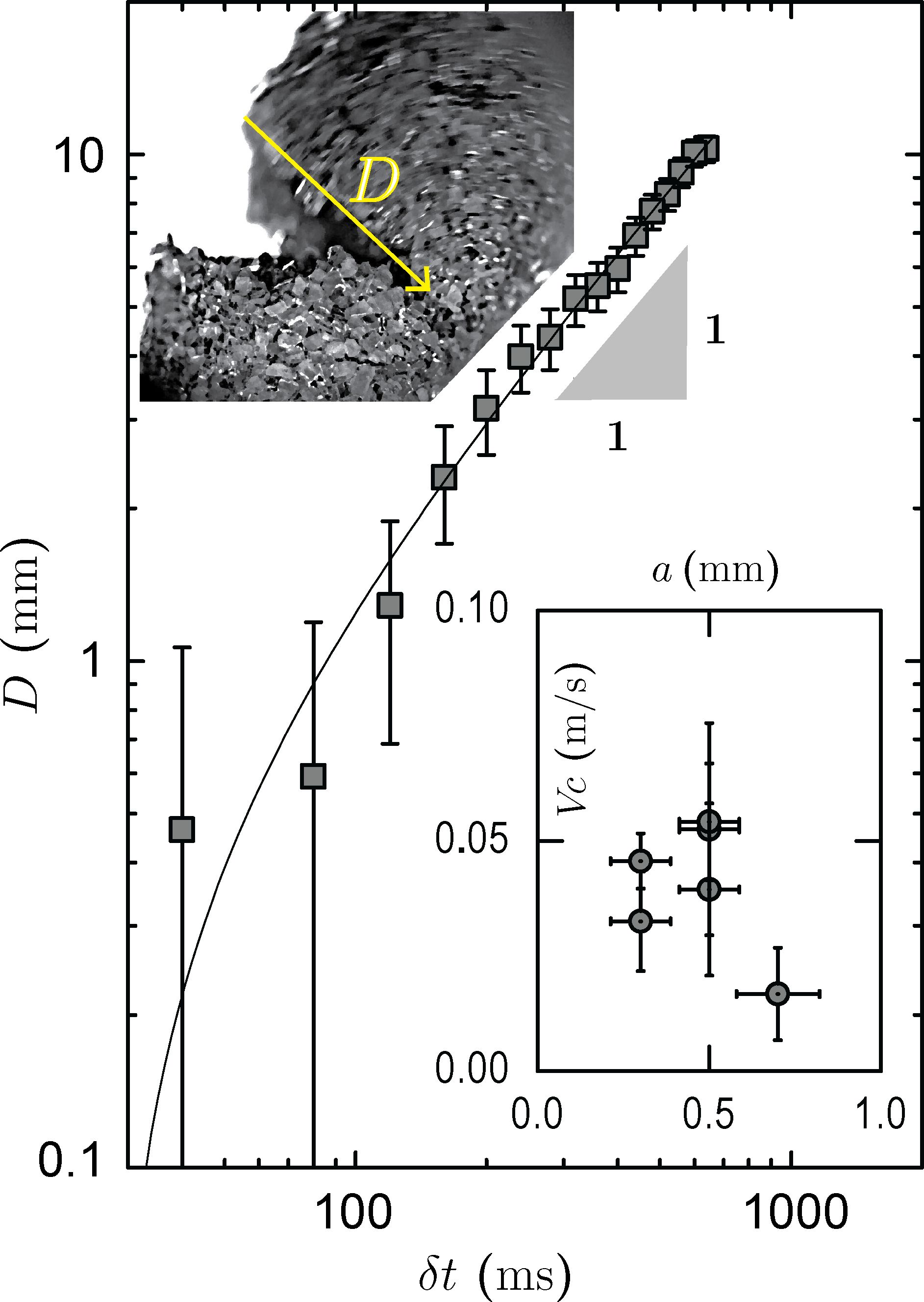}}
		\caption{ Progression of the wedge length $D$ (marked in the inset image) as a function of of the  time $\delta t$  measured onwards from the first observable instance of the wedge opening at the left edge between the falling top and the anchored bottom.  The lower inset shows the variation of the velocity $V_{c}$ at which the breaking front progresses with the grain size $a$. 
		}
		\label{fig:Fig_scaling_C}
	\end{figure}
	
	%%%%%%%%%%%%%%%%%%%%%%%%%%%%%%%%%%%%%%%%%%%%%%%%%%%%%%%%%%%%%% 
	
	The inset image in Fig.\ref{fig:Fig_scaling_C} displays an instantaneous distance $ D $  at the apex of the  \textit{visually identified wedge} between the falling top and the anchored bottom  parts of the column.  This wedge opens up from the trailing (left) edge and progresses towards the advancing (right) edge  of the column.  The variation of this  distance $ D $  with time $ \delta t $ is plotted in Fig.\ref{fig:Fig_scaling_C}, here the  time $\delta t$ is  measured onwards from the first observable instance of the wedge opening at the left edge between the falling top and the anchored bottom. As can be seen, the wedge advances with an average velocity of $  V_c \approx$ \SI{0.02}{m/s}.

	Neglecting the viscous drag, the time required for a non-anchored column to topple 
	can be obtained by equating the toppling torque acting through the center of mass of the falling part of the column and the rate of change of the angular momentum. This `shortest' falling time is about a second which  is the same for the wedge to move across the sample. Accounting for the viscous drag will  increase the estimate of the falling time. Unlike conventional solids which break  by developing  cavities in the bulk, the  interior of the submerged column is made of dry grains that cannot support  an open gap. The gap can  be sustained only if   water drains in and generates a confining \textit{skin} which   prevents   individual particles from falling into it. Hence, the speed at which the  wedge opens sets a lower limit on the  tearing  speed of the \textit{skin}.  The inset of Fig.\ref{fig:Fig_scaling_C} shows that the particle size influences the   speed $ V_c $ at which the apex of the visually identified wedge between the falling top and the anchored bottom moves --  suggesting that the  wedge opening is influenced by   tearing of the \textit{skin}.

	\begin{figure*}[t]
		\centering
		\includegraphics[width=0.99\linewidth]{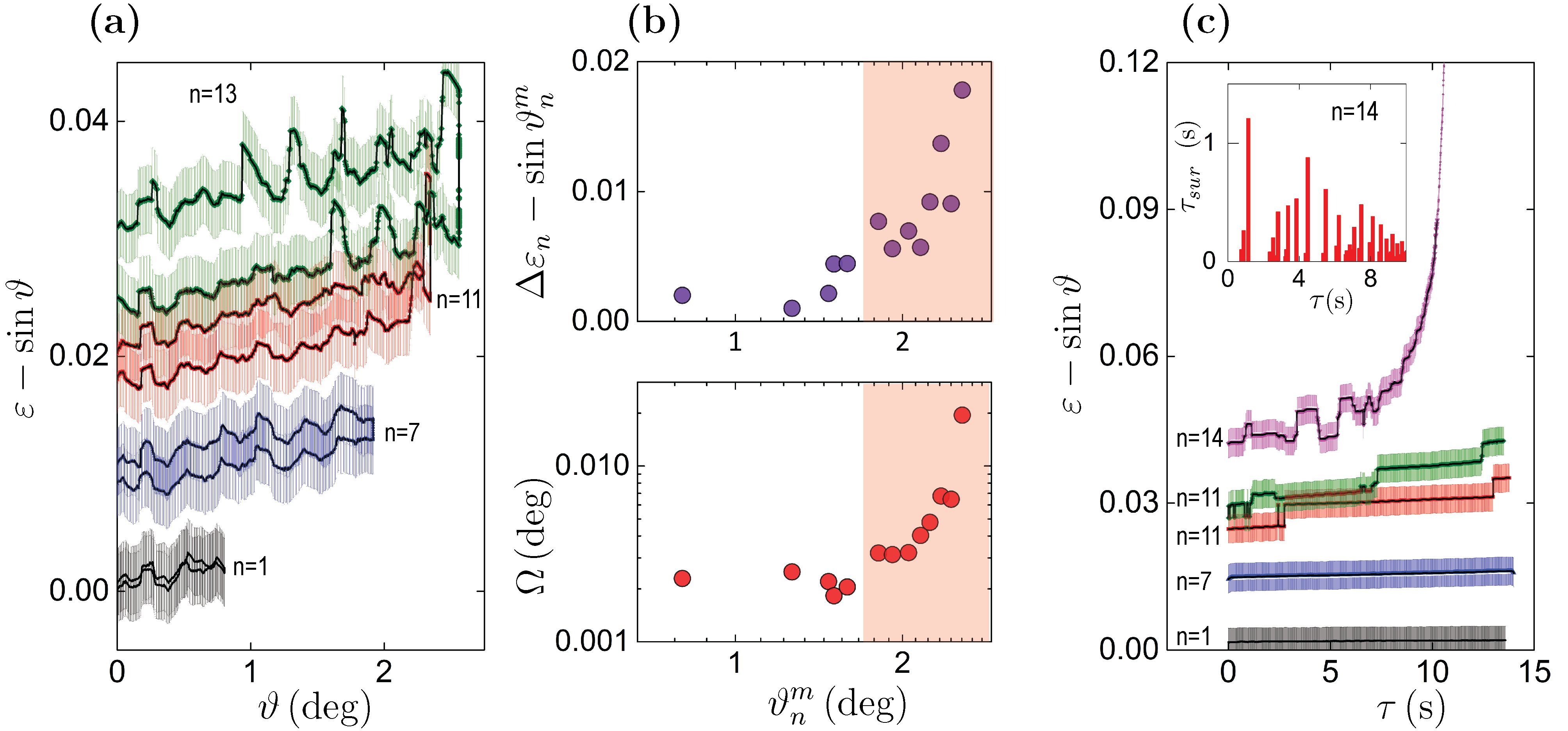}
		\caption{ (a) Variation of $ (\varepsilon - \sin \vartheta ) $ as a function of the tilt angle $\vartheta$ for the entire collection of selective minor-loops, where $ \varepsilon = R/L $. Here, $  R $  is the distance of the mid-point of the top of the column from the point $  (0,L) $  on the $y-$axis. (b) In top and bottom panels, the strain  $(\Delta\varepsilon_n-\sin \vartheta_n^m)$   and total area under the hysteresis loop $\Omega$ are plotted as a function of increasing $\vartheta^{m}_{n}$, respectively. Here, $ \Delta\varepsilon_n = \varepsilon|_n^{\vartheta=0}-\varepsilon|_  n^{\vartheta=\vartheta_n^m } $. The shaded region is dominated by plastic response.
			(c) Variation of $ (\varepsilon - \sin \vartheta ) $ as a function of clamp time $\tau$ for different values of $n$ in the \textit{clamp$- \vartheta $} branch for $ \rm n \,=1,7,11,13\, and\,14 $. The inset shows variation of the survival time $\tau_{surv}$ of a state of the column as a function of $\tau$  for $\mathscr{C}(n=14,\vartheta=2.6^\circ,\tau<10 \, \mbox{s})$. The data shown here are for  a column whose   $L \, = 100 \,\mathrm{mm} $, $a\,=\,500\, \mathrm{\mu m}$,  $d\,= 11\,\mathrm{mm}$.  
		}
		\label{fig:Fig_hist_loops}
	\end{figure*}
	
	%%%%%%%%%%%%%%%%%%%%%%%%%%%%%%%%%%%%%%%%%%%%%%%%%%%%%%%%%%%%%%%%%%%%%%%%%%%%

	\subsection{Effect of stress cycling on the \textit{skin} }
	%%%%%%%%%%%%%%%%%%%%%%%%%%%%%%%%%%%%%%%%%%%%%%%%%%%%%%%%%%%%%%%%%%%%%%%%%%%%  
	%%%%%%%%%%%%%%%%%%  Figure Hystersis Loops:  %%%%%%%%%%%%%%%%%%%%%%%%%%%%%%% 
	%%%%%%%%%%%%%%%%%%%%%%%%%%%%%%%%%%%%%%%%%%%%%%%%%%%%%%%%%%%%%%%%%%%%%%%%%%%%

	Figure  \ref{fig:Fig_curvature} (a)  shows that tilting of the column alters the shape  of the column.  To study the elastic and plastic response of the column, we utilize an incremental stress-cycling protocol; which is the mechanical equivalent of minor hysteresis loop in magnets  ~\cite{barker1983magnetic}. Here, the column is recursively stressed such that the minor loops have three branches: (i) an increase$-  \vartheta $  branch where the tilt angle increases from 0 to $   \vartheta_n^m $  at  a fixed rate (ii) a clamp$-  \vartheta $  branch where the tilt angle is held constant at $   \vartheta_n^m $  for a fixed period of clamping time and (iii) a decrease$-  \vartheta $  branch which corresponds to decreasing the tilt angle from $   \vartheta_n^m $ to 0, with the same rate as in (i). They are accessed sequentially. For each successive cycle, the maximum tilt angle $  \vartheta_{n}^{m}$  increases linearly with $ n$, where $  n $  is the numerical index for the cycle. An instantaneous configuration of the submerged column is referred to as $ \mathscr{C}(n,  \vartheta: \tau)$, where $  \tau $  is the time spent by the system in the clamped state. 
	Fig.\ref{fig:Fig_hist_loops}A shows the cyclic variation in the relative displacement $( \varepsilon - \sin\vartheta ) $ as a function of the tilt angle $   \vartheta $  for selective values of $ n =$ 1, 7, 11 and 13 respectively; here $ \varepsilon=R/L $ and $  R $  is the distance of the mid-point of the top of the column from the point $  (0,L) $  on the $y-$axis (the inset of Fig.\ref{fig:Fig_curvature}(b). 
	For a first few cycles ($n\le 5 $), i.e., for  smaller values of $ \vartheta_n^m$, the \textit{skin} deforms in a reversible manner. We limit our statement of the reversibility: it is entirely possible that while the \textit{skin} may behave in a reversible manner, the dry grains in the interior of the column may not. For larger values of $ \vartheta_n^m$ ($ n>5 $), the column shows noticeable irreversibility and  hence, the hysteresis increases. 
	The parameter $ (\Delta\varepsilon_n-\sin \vartheta_n^m )$ is a measure of the strain developed in each cycle while the parameter $ \Omega $ determines the loop area of each stress cycle; which is a measure of the accumulated hysteresis and dissipation in the system, here $\Delta\varepsilon_n = \varepsilon|_n^{\vartheta=0}-\varepsilon|_  n^{\vartheta=\vartheta_n^m } $.  The top and bottom panels of Fig.\ref{fig:Fig_hist_loops}B show variation of these two parameters as a function of the maximum tilt angle $ \vartheta_n^m $, respectively.

	The following observations can be made from  Fig.\ref{fig:Fig_hist_loops} (a) and (b) : (i) the  elastic regime  of the column extends for  $\vartheta_n^m=1.5^{\circ}$ beyond which the area of the hysteresis loop $ \Omega $ begins to increase abruptly.  This  elastic regime (non-shaded region of Fig. \ref{fig:Fig_hist_loops}(b) can accommodate  strains  of the order of $ 5 \times 10^{-3} $ which is much larger than that observed for dry granular systems.  (ii) In the elastic regime, the  recursive stress cycling  shows smooth   variation in $ (\varepsilon-\sin\vartheta) $ with $ \vartheta $. However, beyond the elastic regime ($ n>5 $ or $ \vartheta_n^m > 1.5^{\circ} $, the shaded region of Fig. \ref{fig:Fig_hist_loops}(b) this variation becomes increasing jagged and clearly shows   discrete `jumps' interspersed with smoothly varying sections. These jumps  are signatures of mechanical instabilities associated with contact line slippages.  They   are related to the stress (tilt)  induced reduction of  the energy barriers  associated with the    underlying pinning potential which  allows the  ambient noise (fluctuations) to  induce  creep like motion of the contact line.   Anomalies of elastic constants associated with the break down of linear elastic response  generates similar jumps in the flow curves of  disordered materials ~\cite{combe2000strain,karmakar2010statistical}.  (iii) The maximum size of the  jump height seen  in the quantity $ (\varepsilon-\sin\vartheta) $   is about $ 0.01 \approx \mathcal{O}(a/L)$, i.e., the contact line on an average moves by a particle scale. Further motion of the contact line is possible only by jumping to the next grain which is restricted by the presence of sharp  edges at the grain corners \cite{dyson1988contact}. 
	(iv) The increase in  $ \Omega $  is caused by the  cumulative effects of these jumps. These jumps progressively generate local  overhangs   along the direction of the  tilt. This  results in  the column  breaking forward unlike a rigid ``falling chimney'' which breaks backward \cite{varieschi2003toy,sutton1936concerning}.

	The effects of stress assisted creep  are  best seen in the  deformation of the column for   the  various clamp $- \vartheta $  branches (here $ \vartheta $ is held constant at $ \vartheta_n^m $, see Fig.\ref{fig:Fig_hist_loops}(c). The creep increases  with increasing tilt angles (forces), e.g., see the  $  n=14 $ branch  in Fig.\ref{fig:Fig_hist_loops}(c).  Even during creep the  column evolves in a similarly punctuated stick-slip manner with jumps in the quantity  $  (\varepsilon-\sin\vartheta) $   being  limited  by the value $ a/L$.
	%In the stick-phase the state of the column appears to be stationary while it evolves in the slip-phase.
	
	As  the occurrences of creep increase, the  
	survival$-$time $  \tau_{surv} $  of  a stationary stick phase decreases; see the  inset of Fig.\ref{fig:Fig_hist_loops}(c). The jumps themselves are abrupt in the experimental time scale: they occur over a period less than \SI{10}{ms}, an order of magnitude shorter than the measured shortest survival time (\SI{100}{ms}). The breaking of the column at the macroscopic scale (see Fig.\ref{fig:Fig_scaling_C}) is visible only for values of $ \tau $ greater than \SI{10}{s}.  This region shows a rapid growth of $ (\varepsilon-\sin\vartheta) $, its initiation is marked by an arrow in Fig.\ref{fig:Fig_hist_loops}(c).

	\section{Brittle to ductile to a single particle disintegration: Tuning the failure mode }
\begin{figure}[t]
		\centering {\includegraphics[width=1\linewidth]{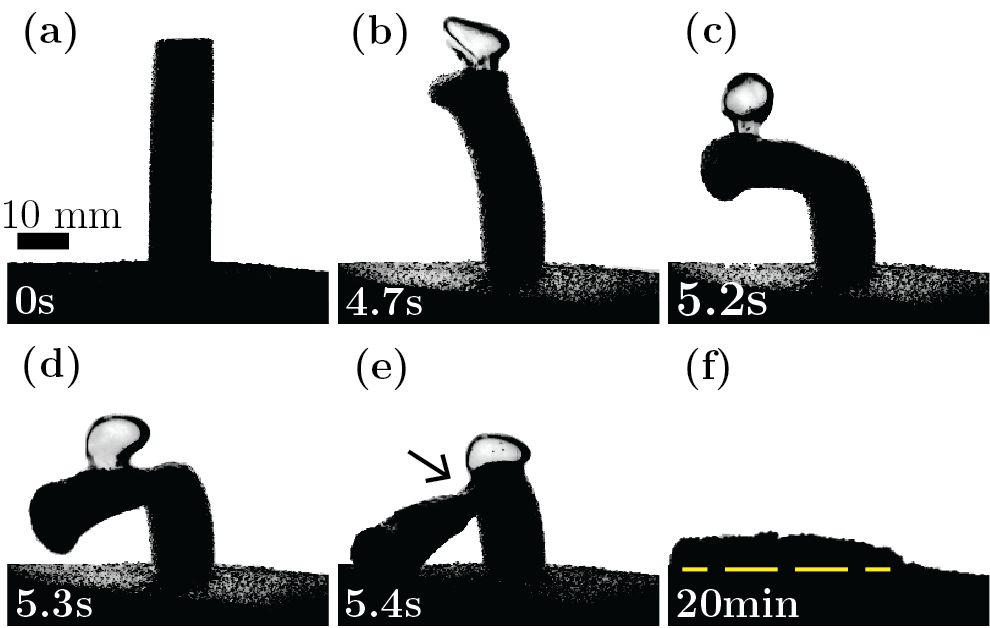}}
		\caption{ A sequence of images: (a)$\rightarrow $(e) shows an upright sand column ($d$ = \SI{14}{mm} and $L$ = \SI{50}{mm}) which fails in a ductile manner. This mode of failure is initiated  by  degassing the column and achieved by maintaining a low vacuum state over the free surface of water. The degassing proceeds by emanating bubbles from the column, which reduces the pressure inside the column and thereby forcing conformational changes on the \textit{skin}. (f) The column looses its integrity by the slightest mechanical disturbance and the structure disintegrates into a heap of particles under water.  }
		\label{fig:Fig_vcaccum_column}
	\end{figure}

	These results imply that the \textit{skin}, comprising of the air-water-grain interface on the surface layer of grains, is primarily responsible for the mechanical response of the entire system and any modification of the \textit{skin} will modify the behaviour of the system. One striking modification is shown in Fig.\ref{fig:Fig_vcaccum_column} where we create a partial vacuum  over the water surface, so that the trapped air within the column, one key component of the \textit{skin}, escapes out from the system as bubbles. The time-lapsed images in Fig.\ref{fig:Fig_vcaccum_column} (a) $\rightarrow$ (e)  demonstrate that here too the failure is initiated by bending at a finite critical height. But, in contrast to the single tear shown above, the column shows large-scale plastic deformations. Its structural integrity is extended for values of $ \psi $ greater than $ 90^{\circ} $ by forming a deformed neck, i.e., a narrowing of the column's diameter, instead of forming a tear or a crack as above. Here, the failure proceeds via this necking instability whose spatial location is marked by an arrow in Fig.\ref{fig:Fig_vcaccum_column}(e). Such large-scale (many-grains wide, in this case) plastic flow and necking instabilities are typical of ductile mode of failure in solids ~\cite{antolovich2014plastic}, in complete contrast with the brittle failure described above. This change in failure-mode is brought about by the loss of entrapped air causing weaker cohesion in the column that results in conformational changes of a softer \textit{skin} and finally leads to the ductile failure of the column.
	
	For completeness, we note that a third, and limiting, mode of failure occurs if degassing continues for a long time. In the absence of air, the \textit{skin} begins to lose its strength (see the lustre-less lump in Fig.\ref{fig:intro}(c). The weak \textit{skin} eventually fails to sustain the inward hydrostatic pressure and slumps under the slightest mechanical disturbance. The water quickly drains in as the structure disintegrates into a heap of particles, seen in both Fig.\ref{fig:intro}(d) and Fig.\ref{fig:Fig_vcaccum_column}(f); the hydrophobic sand then behaves like its hydrophilic counterpart (right panel of Fig.\ref{fig:Fig_modulation}(b). The  removal of trapped air from the column  demonstrates the existence of an underlying  dynamical transition via which the  failure-mode transforms from being brittle to being ductile, both of which are collective (multi-particles) in nature, and finally to a total disintegration of the \textit{skin} that represents failure at the single-particle level. This is analogous to dynamical transitions from collective response to single-particle response in a wide class of systems ~\cite{PhysRevLett.105.154301}.
	
	%%%%%%%%%%%%%%%%%%%%%%%%%%%%%%%%%%%%%%%%%%%%%%%%%%%%%%%%%%%%%%%%%%%%%%%% 
	%%%%%%%%%%%%%%%%%%  Figure vacuum column:  %%%%%%%%%%%%%%%%%%%%%%%%%%%%%%% 
	%%%%%%%%%%%%%%%%%%%%%%%%%%%%%%%%%%%%%%%%%%%%%%%%%%%%%%%%%%%%%%%%%%%%%%%% 
	
	%%%%%%%%%%%%%%%%%%%%%%%%%%%%%%%%%%%%%%%%%%%%%%%%%%%%%%%%%%%%%% 

	\section{Conclusion}

	This paper provides a detailed study of mechanical properties of underwater granular structures made of hydrophobic sand, where the self-generated cohesive \textit{skin} on the boundary  of the structure encapsulates the dry grains inside it from the surrounding medium water. In our experimental results, three distinct length scales of the system are found. (i) The scale of the hydrophobic patches $\approx \mathscr{O} (1 \, \mu\mbox{m}) $ influences the strength of the pinning of air-water interface on the grain surfaces. Depinning of this interface produces macroscopically observable plastic deformations of the contact lines around (ii) the grain-scale $ \approx \mathscr{O} (10^2 \, \mu \mbox{m}) $ of the sand particles. (iii) At the system scale $ \approx \mathscr{O} (10^4 \, \mu\mbox{m}) $, the compressive traction forces due to the column's own weight drives the wrinkling of the \textit{skin}. The regions of large curvature of the structure are the seeds from where the system-sized failure modes nucleate. By partially removing the trapped air from this structure, we also see  that the collective failure can be tuned from brittle to ductile. A more complete removal of air causes the \textit{skin} to crumble completely into individual non-cohesive grains inside water. These experimental findings imply the existence of a tunable dynamical transition between a collective and an individual (single-grain) mode of  failure in this system. We expect that these new results will help in engineering the granular encapsulation with desired material-properties in a variety of applications and, at the same time, it provides a deeper insight of the multi-scale mechanics, generic to  granular materials.

	We acknowledge discussions with A.  Ghatak, M.  Tirumkudulu, I.  Sharma and M. Bandi. We especially thank the anonymous referees for many critical and valuable comments on the manuscript.

	%\bibliography{main_ref_resubmit}
	%%%%%%%%%%%%%%%%%%%%%%%%%%%%%%%%%%%%%%%%%%%%%%%%%%%%%%%%%%%%%%
	%      supplementary figures
	%%%%%%%%%%%%%%%%%%%%%%%%%%%%%%%%%%%%%%%%%%%%%%%%%%%%%%%%%%%%%%  
	%\input{supp_file}
	%%%%%%%%%%%%%%%%%%%%%%%%%%%%%%%%%%%%%%%%%%%%%%%%%%%%%%%%%%%%%%
	%merlin.mbs apsrev4-1.bst 2010-07-25 4.21a (PWD, AO, DPC) hacked
	%Control: key (0)
	%Control: author (8) initials jnrlst
	%Control: editor formatted (1) identically to author
	%Control: production of article title (-1) disabled
	%Control: page (0) single
	%Control: year (1) truncated
	%Control: production of eprint (0) enabled
	%

	\FloatBarrier

	%\begin{linenumbers}
	
\end{document}